\documentclass[aps,prl,twocolumn,superscriptaddress]{revtex4-1}
\usepackage{graphicx}
\usepackage[outdir=figures./]{epstopdf}

\usepackage{hyperref}
\begin{document}


\title{Rattling mode and symmetry lowering resulting from instability of B$_{12}$-molecule in LuB$_{12}$}


\author{N.\/Sluchanko}
\email[]{nes@lt.gpi.ru}
\affiliation{Prokhorov General Physics Institute, Russian Academy of Sciences, 38 Vavilov Str., 119991 Moscow,  Russia}
\affiliation{National University of Science and Technology (MISiS), 119049 Moscow, Russia}

\author{A.\/Bogach}
\affiliation{Prokhorov General Physics Institute, Russian Academy of Sciences, 38 Vavilov Str., 119991 Moscow,  Russia}

\author{N.\/Bolotina}
\affiliation{Shubnikov Institute of Crystallography of Federal Scientific Research Centre “Crystallography and Photonics” of Russian Academy of Sciences, 59 Leninskii Ave., 119333 Moscow,  Russia	}

\author{V.\/Glushkov}
\affiliation{Prokhorov General Physics Institute, Russian Academy of Sciences, 38 Vavilov Str., 119991 Moscow,  Russia}
\affiliation{National University of Science and Technology (MISiS), 119049 Moscow, Russia}

\author{S.\/Demishev}
\affiliation{Prokhorov General Physics Institute, Russian Academy of Sciences, 38 Vavilov Str., 119991 Moscow,  Russia}
\affiliation{Moscow Institute of Physics and Technology (State University), 9 Institutskiy Per., 141700 Dolgoprudny,  Russia}

\author{A.\/Dudka}
\affiliation{Shubnikov Institute of Crystallography of Federal Scientific Research Centre “Crystallography and Photonics” of Russian Academy of Sciences, 59 Leninskii Ave., 119333 Moscow,  Russia	}

\author{V.\/Krasnorussky}
\affiliation{Prokhorov General Physics Institute, Russian Academy of Sciences, 38 Vavilov Str., 119991 Moscow,  Russia}

\author{O.\/Khrykina}
\affiliation{Shubnikov Institute of Crystallography of Federal Scientific Research Centre “Crystallography and Photonics” of Russian Academy of Sciences, 59 Leninskii Ave., 119333 Moscow,  Russia	}

\author{K.\/Krasikov}
\affiliation{Moscow Institute of Physics and Technology (State University), 9 Institutskiy Per., 141700 Dolgoprudny,  Russia}

\author{V.\/Mironov}
\affiliation{Shubnikov Institute of Crystallography of Federal Scientific Research Centre “Crystallography and Photonics” of Russian Academy of Sciences, 59 Leninskii Ave., 119333 Moscow,  Russia	}

\author{V.\/Filipov}
\affiliation{Frantsevich Institute for Problems of Materials Science, National Academy of Sciences of Ukraine, 3 Krzhyzhanovsky Str., 03680 Kiev,  Ukraine}

\author{N.\/Shitsevalova}
\affiliation{Frantsevich Institute for Problems of Materials Science, National Academy of Sciences of Ukraine, 3 Krzhyzhanovsky Str., 03680 Kiev,  Ukraine}

\date{July 10, 2017}

\begin{abstract}
The dodecaboride LuB$_{12}$ with cage-glass state and rattling modes has been studied to clarify the nature of the large amplitude vibrations of Lu ions. Discovered anisotropy of charge transport in conjunction with distortions of the conventional \textit{fcc} symmetry of the crystal lattice may be attributed to coherent motion of Lu ions along singular direction in the lattice. Arguments are presented in favor of cooperative dynamic Jahn-Teller effect in the boron sublattice to be the reason of the rattling mode, lattice distortion and formation of the filamentary structure of the conductive channels.
\begin{description}

	\item[PACS numbers: 61.66.Fn, 72.15.Gd]

\end{description}
\end{abstract}

\pacs{Valid PACS appear here }

\maketitle

\textit{Introduction.} -- Rattling compounds have attracted wide interest in the last few decades due to their intriguing and exotic physical properties. Rattling effect (a large amplitude vibration of an atom in an oversized atomic cage) can be responsible for an extremely low thermal conductivity and high thermoelectric efficiency in various cage compounds, such as filled skutterudites \cite{keppens98, *nakai2008}, $\beta$-pyrochlore oxides \cite{yoshida2007}, clathrates \cite{nolas98,*paschen14}, $R$T$_2$Zn$_{20}$ ($R$= Pr, La;T= Ir, Ru) \cite{asaki12}, quadruple perovskites \cite{akizuki15}, higher borides RB$_{12}$ \cite{sluchanko09, *alekseev14}, etc. These local excitations may also affect the electronic properties of solids due to strong electron-phonon coupling \cite{sluchanko09, *alekseev14}.  The nature of rattling motion and the mechanisms of its influence on the unusual properties and exotic ground states is still under question.

\begin{figure}
	\includegraphics{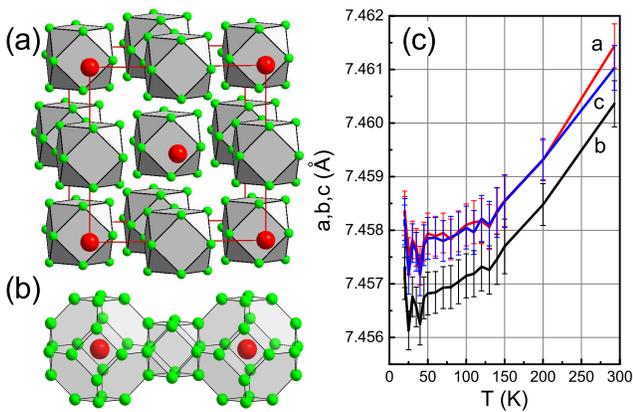}
	\caption{\label{fig:fig1} (Color online) (a) a NaCl-like unit cell of LuB$_{12}$, with Lu (red spheres) as Na and B$_{12}$ clusters (green spheres) as Cl. (b) two large B$_{24}$ polyhedra centred by Lu atoms and a smaller B$_{12}$ cubooctahedron between them. (c) The temperature dependence of the lattice parameters $ a, b \text{ and } c $.}
\end{figure}

Among rattling compounds the rare earth and transition metal dodecaborides $R$B$_{12}$ ($ R $ = Y, Zr, Tb-Lu) represent the simplest, model objects with face-centered cubic ({\textit {fcc}}) NaCl-type crystal lattice  (space group \textit {Fm}$ \overline{3} $\textit{m}-O5h) built by B$_{12}$ cubooctahedra and metal atoms $ R $ centered in the large cuboctahedral cages B$_{24}$ formed by six neighboring B$_{12}$ units (Fig.~\ref{fig:fig1}a,b). 
Strong covalent bonds between boron atoms (both within B$_{12}$ units and between them) form a rigid boron framework (Fig.~\ref{fig:fig1}b), which changes insignificantly in the $R$B$_{12}$ family. 
Large difference between the size of B$_{24}$ cage (r$(\text{B}_{24}) \sim 1.1-1.15 $\AA{} \cite{alekseev08}), and the radius of metallic ions ($0.8-0.97$\AA{}) leads to a formation of loosely bound state of the heavy ion in the rigid  B$_{24}$ cage, resulting in low frequency ($ 14-18 $ meV \cite{rybina}) dispersion-less (Einstein-like, rattling) vibrations in the phonon spectrum of dodecaborides. 
The electron deficiency in the boron lattice is compensated by transfer of two valence electrons ($6s^2$) from each $R$ atom to B$_{12}$ cluster, while the third valence electron (5$d^1$) in $R$B$_{12}$ enters the conduction band. So, all rare earth  dodecaborides are good metals, in which conduction band at the Fermi level is mainly contributed from $5d$-states with a small admixture of B 2$p$-electrons \cite{heinicke, *jager}. The only exception is a narrow-gap semiconductor YbB$_{12}$ with exotic insulating ground state, which is observed in the regime of strong charge and spin fluctuations. Besides, YbB$_{12}$ undergoes a metal-insulator transition at $T^*\sim60$ K \cite{sluchanko09, alekseev14, alekseev08,mignot}. Thus, these $R$B$_{12}$ rattling compounds cover a variety of ground states and the regimes of charge transport, that ranges from the unusual superconductivity in LuB$_{12}$ and ZrB$_{12}$ \cite{alekseev08} and Kondo insulator \cite{sluchanko09, alekseev14, alekseev08,mignot} or topological crystalline insulator behaviour \cite{weng} predicted for YbB$_{12}$ to the complicated magnetic ordering and peculiar incommensurate magnetic structures detected  for TbB$_{12}$-TmB$_{12}$ antiferromagnets \cite{alekseev08}. 

It is particularly important that there are some experimental evidences for the local structural distortions in the cubic lattice of $R$B$_{12}$ at low temperatures. It has been recently concluded \cite{sluchanko11} that these dodecaborides $R$B$_{12}$  tend to form a cage-glass phase through the order-disorder phase transition (at $T^* \sim 60$ K in LuB$_{12}$),  that results in freezing of $R$ ions in the random positions inside the B$_{24}$  cages. Besides, x-ray diffraction studies \cite{bolotina} discovered a significant tetragonal distortion of the atomic structure of LuB$_{12}$  in the vicinity of this phase transition.

Given that LuB$_{12}$ is a reference compound for the family of $R$B$_{12}$, it is important ($i$) to investigate an interplay between the rattling effect and features of crystalline and electronic structure and ($ii$) to shed more light on the nature of rattling in this non-magnetic dodecaboride. 
Analysing the influence of rattling mode on the transport and structural properties of LuB$_{12}$ we studied in detail the changes of crystal structure and the anisotropy of charge carriers scattering developed at low temperatures in the cage-glass state of LuB$_{12}$. 
Then, searching for the mechanism governing the large amplitude vibration of $R$ ion we present here the results of quantum chemical calculations and geometry optimizations for negatively charged $[\text{B}_{12}]^{2-}$ cluster.  Our results argue in favour of a cooperative dynamic Jahn-Teller effect in the boron sub-lattice as a possible mechanism responsible for the rattling modes, structural distortions and charge transport anisotropy in LuB$_{12}$.

\begin{figure*}
	\includegraphics{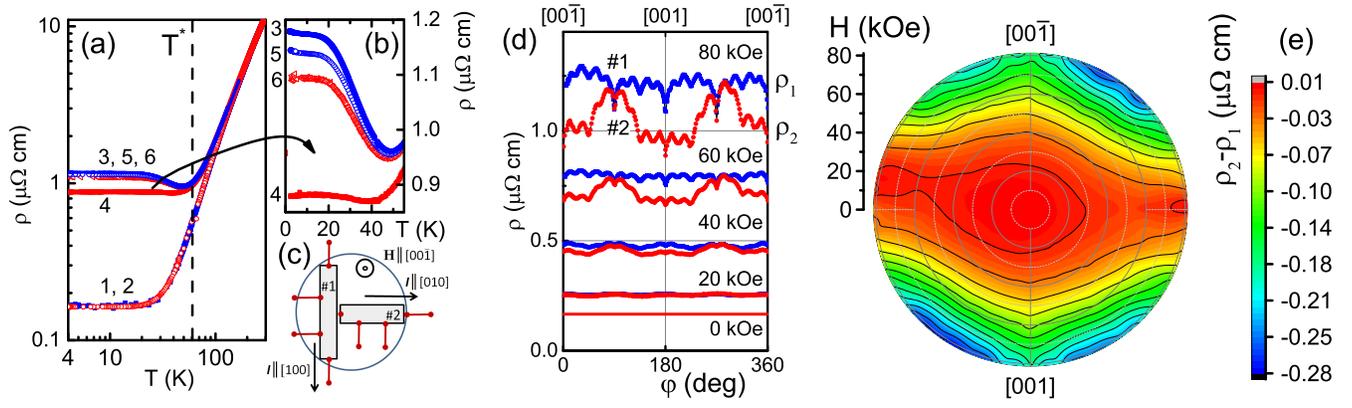}
	\caption{\label{fig:fig3} (Color online) (a,b) Temperature dependences of the resistivity $ \rho(T) $ for LuB$_{12}$  crystals \#1($\mathbf {I} ||$[100]) and \#2 ($\mathbf {I} ||$[010]) are shown in the absence of an external magnetic field ($ H = 0 $, curves 1,2 corespondingly) and in steady magnetic field of 80 kOe directed along the crystal axes [00$\overline{1}$] (curve 3 for \#1, curve 4 for \#2)), [0$\overline{1}$0] (curve 5 for \#1) and [$\overline{1}$00] (curve 6 for \#2). (c)Two [100] and [010] -elongated rectangular samples \#1 and \#2 with equally oriented (100), (010) and (001) faces were cut out from one single-crystalline disk of LuB$_{12}$. Angular dependences of $\rho_1(\varphi)$ and $\rho_2(\varphi)$ obtained rotating correspondingly the crystals \#1 and \#2 around their current axes in various magnetic fields up to 80 kOe at temperatures 2 - 4.2K. (e) The anisotropy of magnetoresistance $\rho_2(\varphi) - \rho_1(\varphi) = f(\varphi, H)$ presented in the polar coordinates.}
\end{figure*}

\textit{Experimental results.} -- Details of experiment are given in Supplementary materials \cite{[See] [{for more information: (Details of experiments, including single crystals growth and characterization (precise x-ray diffraction reflexes [Fig. 1S] and de Haas-van Alphen effect [Fig. 2S]), the schema of magnetoresistance measurements [Fig.S3]. Illustration of a symmetry lowering in LuB$_{12}$:  x-ray data are collected at T = 140 K and difference Fourier maps are obtained when structural symmetry is described using \textit {Fm}$ \overline{3} $\textit{m} (a-c) and F\textit{mmm} (d-f) groups [Fig. S4]. In depth information on magnetoresistance as obtained for samples \#1 and \#2 [Figs.S5]. Results of quantum chemical calculations for electrically neutral $[\text{B}_{12}]^0$ and negatively charged clusters $[\text{B}_{12}]^{n-}$ ($ n = 1 - 4 $) with detailed discussion)}]suppl}. 
Precise x-ray diffraction reflexes  and quantum oscillations of magnetization (see Figs. in \cite{suppl}), testify to the high quality of the crystals. 
The [100]- and [010]- elongated rectangular samples (\#1 and \#2 hereafter) with equally oriented (100), (010) and (001) faces were cut from one ingot of LuB$_{12}$ (Fig.\ref{fig:fig3}c). 
The resistivity $\rho$(T) curves for LuB$_{12}$ crystals \#1 (measuring current $\mathbf{I}||$[100])and \#2 ($\mathbf{I}||$[010]), are shown (Fig.\ref{fig:fig3}a) in the absence of an external magnetic field (H = 0, curves 1 and 2 correspondingly) and in steady magnetic field of 80 kOe directed along the crystal axes [00$\overline{1}$] (curve 3 for \#1, curve 4 for \#2), [0$\overline{1}$0] (curve 5 for \#1) and [$\overline{1}$00] (curve 6 for \#1 and 7 for \#2). 
A significant anisotropy of the transverse magnetoresistance ($\sim 20$\%)  is observed below $T^* \sim$  60 K at $H=80  $ kOe, especially for the crystal \#2 (see curves 4 and 6 in Fig.\ref{fig:fig3}a, \ref{fig:fig3}b for comparison), in spite of the fact that $\mathbf{H}$-directions [00$\overline{1}$] and [$\overline{1}$00] are symmetry-equivalent in cubic crystals. 
 In addition, a minimum is observed below $T^*$ on the dependences $\rho(T, H=80\, \text{kOe})$ and the temperature lowering is accompanied with a growth of resistivity (see Fig.\ref{fig:fig3}b). Such behaviour cannot be attributed to the Kondo scattering of charge carriers, because it is accompanied with a large positive magnetoresistance in these LuB$_{12}$ crystals.
 Note that the resistivity $\rho(T, H=80\, \text{kOe})$ minimum below  $T^*$ is accompanied with a maximum in the vicinity of 60 K on the R$_H(T)$ curves for $\mathbf{I} ||$[110], which was detected earlier both in  Lu$^N$B$_{12}$ with various isotopic composition in boron ($ N $-natural, 10 and 11) \cite{sluchanko10} and in substitutional solid solutions Zr$_{1-x}$Lu$_x$B$_{12}$ \cite{sluchanko16} and this anomaly of Hall effect was associated with the transition to the cage-glass state.

To clarify the nature of the anisotropy of the transverse magnetoresistance at  low temperatures, angular dependences $\rho(\varphi)$ were measured at  2\,-\,4.2 K in the magnetic fields up to 80 kOe by rotating the crystals \#1 and \#2 around their current axes. Families of curves $\rho_1(\varphi)$ for $\mathbf{I} ||$[100] (sample \#1) and $\rho_2(\varphi)$ for $\mathbf{I} ||$[010] (sample \#2) are shown in Fig.\ref{fig:fig3}d. 
Evidently, the angular dependences of  $\rho_2(\varphi)$ do not correspond to those expected for a cubic structure. As mentioned above, the great difference is recorded for $ \mathbf{H}||[00\overline{1}] (\varphi = 0^\circ)$ and $ \mathbf{H}||[\overline{1}00] (\varphi = 90^\circ) $, although these directions are equivalent in cubic crystals. On the contrary, the sample \#1 gives a set of $\rho_1(\varphi)$ singularities corresponding to practically equivalent directions $[00\overline{1}], [0\overline{1}0]$ (see Fig.\ref{fig:fig3}d and \cite{suppl}).
 The anisotropy of magnetoresistance is clearly resolved in the polar coordinates in Fig.\ref{fig:fig3}e, where the difference $\rho_2(\varphi) - \rho_1(\varphi) = f(\varphi, H)$ is presented in the coloured picture.
 It is clearly discerned from Fig.\ref{fig:fig3}e that conduction channels, which are transverse to the plane (001), appear in the LuB$_{12}$ matrix.

\textit{Crystal structure.} -- In accordance with arguments presented in \cite{bolotina}, the \textit{fcc} symmetry of LuB$_{12}$ is notably distorted at low temperatures. Conventional software for the structure analysis results in calculated electron density (ED) replicating the symmetry of structure model if even the symmetry of real ED is distorted. To provide the ability to derive probable violations of symmetry from difference Fourier maps, the crystal structure must be analysed using a less symmetrical model (see \cite{suppl} for details). Unit-cell values of LuB$_{12}$ were refined without any symmetry restriction over the temperature range 20-300 K. Very small ($\sim 0.001 $ \AA{}, see Fig.\ref{fig:fig1}c) but steady difference of the lattice constants $ a, b, c $ was revealed, which was not accompanied by steady deviation of angular parameters from $90^\circ$.
Cubic metric of the unit cell was kept, since revealed differences in the unit-cell values $ b \neq a \cong c $ (Fig.\ref{fig:fig1}c) was too small to influence on a result of the structure refinement. Thermal vibration of the Lu ions was described by an isotropic parameter in order that assumed anisotropy of residual ED near the Lu site could reveal itself most clearly.

\begin{figure*}
	\includegraphics[width=7.1in, keepaspectratio]{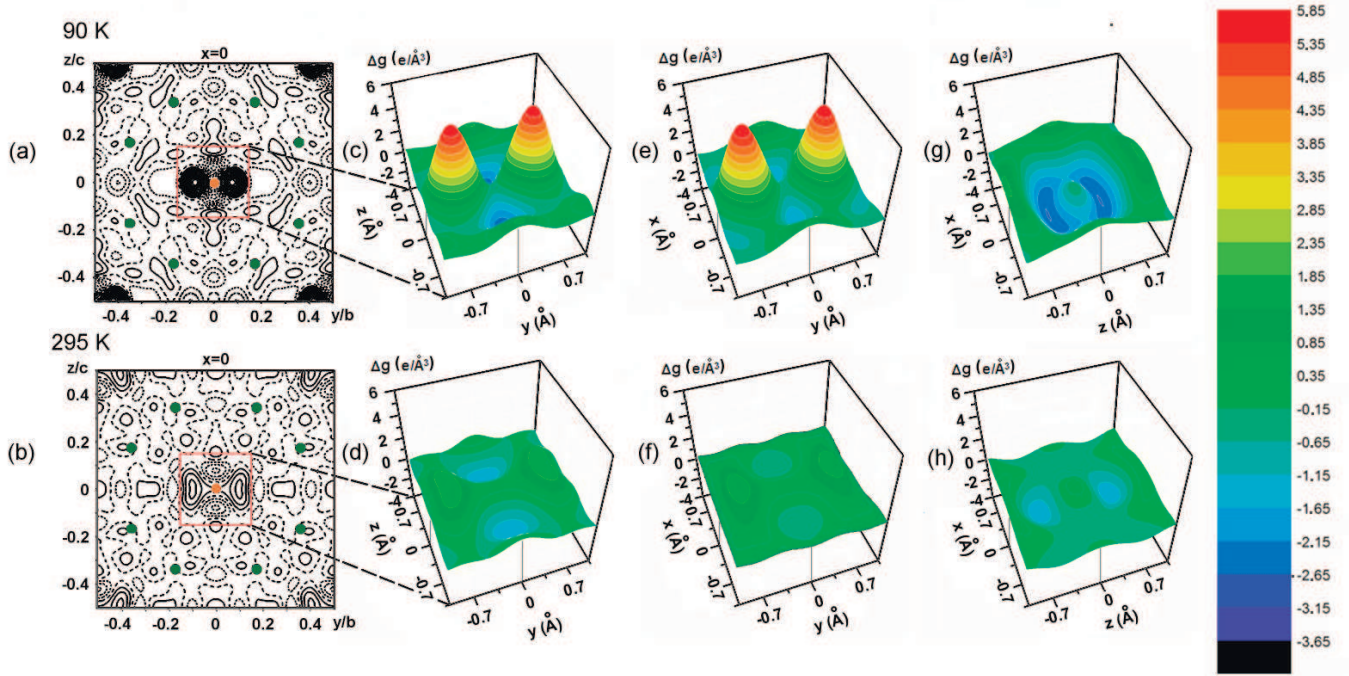}
	\caption{\label{fig:fig2} (Color online) Difference Fourier maps (residual ED $ \Delta g, e$/\AA{}$^3$) in the $ x=0 $ face of the LuB$ _{12} $ unit cell at 90 (a) and 295 K (b), respectively. Red circle is the Lu-site; green circles are B-sites. The (c, d), (e, f) and (g, h) panels are relief drawings of difference Fourier maps made in the vicinity of the Lu-ion, in the $ x = 0 $, $ z = 0 $ and $ y = 0 $ faces of the unit cell, whereas first and second rows of overall picture correspond to temperatures 90 K and 295 K, respectively.}
\end{figure*}

Difference Fourier maps were built based on the results of the \textit{Fmmm} structure refinement at  temperatures 90 and 295 K  (Fig.~\ref{fig:fig2}, see \cite{suppl} for details). Each Lu site in the cubic \textit {Fm}$ \overline{3} $\textit{m} structure lies at the intersection of three 4-fold axes perpendicular to three faces of the cubic cell. As may be seen (Fig.\ref{fig:fig2}a,b), residual ED distribution near Lu$ (0,\frac{1}{2},\frac{1}{2}) $ in the x = 0 face is not bound by a 4-fold axis even at room temperature, what does not actually agree with \textit {Fm}$ \overline{3} $\textit{m}.

\begin{figure*}
	\includegraphics [width=7.1in, keepaspectratio]{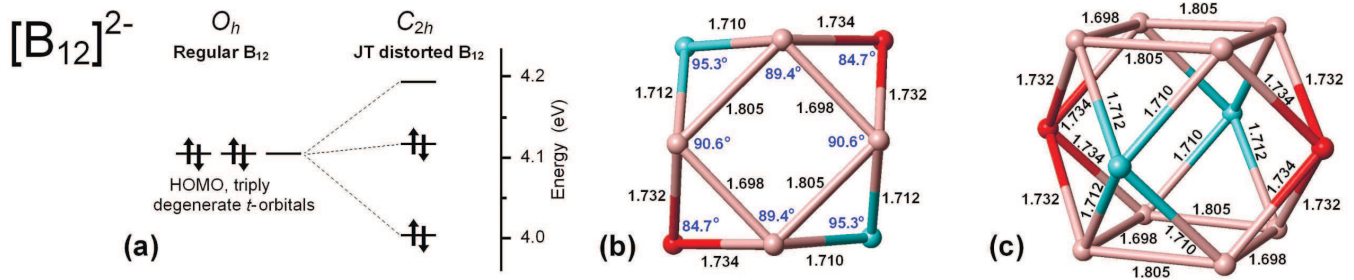}
	\caption{\label{fig:fig4} (Color online) (a) JT splitting of triply degenerate highest occupied molecular orbital with four electrons is indicated. Molecular structure of isolated cluster [B$_{12}$]$^{2-}$ in the local JT energy minimum as obtained from DFT geometry optimization calculations \cite{suppl} (projection (b) and perspective view (c)). Principal atomic distances (\AA{}) and bond angles are indicated. The structure exhibits the $C_{2h}$ symmetry. There are three groups of equivalent (symmetry related) boron atoms shown as pink, red, and blue balls.}
\end{figure*}

Relief fragments of difference Fourier maps at $ x = 0 $ (Fig.~\ref{fig:fig2}c,d), $ z = 0 $ (Fig.~\ref{fig:fig2}e,f), $ y = 0 $ (Fig.~\ref{fig:fig2}g,h) are presented in three next columns of Fig.\ref{fig:fig2} for better visualization of residual ED distribution near the Lu site. Most notable maxima of residual ED are observed along the $y$-axis, along which the smaller period $b\neq a \cong c$ is detected (Fig.\ref{fig:fig1}). Amplitudes of residual peaks of ED increase significantly with temperature decrease being located $ \sim 0.4-0.7 $\AA{} away from a central Lu site. Taking temperature behaviour of the unit-cell values into account, the character of structural distortion is closer to tetragonal. Unique $y$-axis is not an ideal 4-fold axis 
but can be considered as such a symmetry element to a certain approximation \cite{suppl}.

\textit{Discussion.} -- To understand possible reasons for (\textit{i}) the symmetry deviation of the LuB$_{12}$ crystal from the cubic one, (\textit{ii}) the significant anisotropy of the transverse magnetoresistance and the associated non-equivalence of the directions $\mathbf{H}||[00\overline{1}]$ and $\mathbf{H}||[\overline{1}00]$ in the charge carriers scattering (Figure \ref{fig:fig3}d-e, \cite{suppl}), and (\textit{iii}) the growth of resistance with the temperature lowering in a steady magnetic field (Fig.~\ref{fig:fig3}b), we discuss herein a possible scenario associated with the Jahn-Teller (JT) effect. More specifically, because of triple orbital degeneracy of the ground electronic state, the B$_{12}$  molecules are JT-active and thus their structure is labile due to JT-distortions. In this case, upon decreasing temperature, some intrinsic structural defects (such as boron vacancies and mixed $^{10}$B - $^{11}$B isotope composition of B$_{12}$ molecules) can lift degeneracy due to symmetry lowering and may cause an electronic phase transition associated with JT structural instability. Similar phase transformations produced by JT-effect of  B$_{12}$ icosahedra were earlier observed in higher borides \cite{franz}.

In order to establish the amplitude and type of the JT distortions in B$_{12}$ cuboctahedra, we performed quantum chemical calculations and geometry optimizations for charged [B$_{12}$]$^{2-} $ cluster, whose doubly negative charge state is regarded as the most relevant in $R$B$_{12}$ compounds (see \cite{suppl}). The structure is shown in Fig.~\ref{fig:fig4}. Interestingly, the actual symmetry of the JT distorted cluster ($C_{2h}$) is lower than the symmetry of tetragonal ($D_{4h}$) and trigonal ($D_{3d}$) JT minima expected from the JT theory \cite{bersuker}. The amplitude of the JT distortions of isolated B$_{12}$ clusters is rather pronounced as the bond lengths and bonding angles can vary by $\sim 0.1 $\AA{} and $\sim 5^\circ$, respectively (Fig.~\ref{fig:fig4} and  \cite{suppl}).

These results provide evidence that the JT structural liability of B$_{12}$ clusters should play an important role in the microscopic mechanism of lattice distortions of LuB$_{12}$ at low temperatures. Taking into account that B$_{12}$ clusters form an extended 3D covalent network, being connected by B-B covalent bonds, one can suggest that structural JT liability should retain and reinforce in the boron sublattice of the dodecaborides. The reinforcement due to cooperative dynamic JT effect manifested both in static and dynamic lattice properties may be considered as the cause of large amplitude displacements of Lu atoms in oversized B$_{24}$ cages, resulting in quasi-tetragonal distortions of the \textit{fcc} lattice and the anisotropy of magnetoresistance in LuB$_{12}$ at low temperatures. 
Therefore, the rattling modes can be attributed to quantum motion of the Lu ions in the double-well potentials (DWPs) with the minima displaced from each other on the distance of $0.4 - 0.7 $\AA{} along the $y$-axis (see Figs. \ref{fig:fig2}c and \ref{fig:fig2}e). 
As a sequence of these large amplitude vibrations of heavy ions the $ 5d-2p $ hybridization of electron states along the unique $y$-axis changes dramatically resulting in formation of conductive channels (dynamic stripes \cite{REZNIK201275}) with a strong charge carrier scattering on the filamentary structure. Thus, the proposed scenario may be used to explain naturally the anisotropy of the charge transport in LuB$_{12}$.

The barrier height $ \Delta E = 51 - 97 K $ in DWP is detected \cite{sluchanko14} for LuB$_{12}$ with different numbers of boron vacancies and Zr impurities. Concentration of Lu ions displaced from centres in B$_{24}$ cubooctahedra is estimated to be 3 - 8 at.\%. This agrees well with the 3.6 at.\% concentration of the off-site Lu ions obtained from the EXAFS measurements at low temperatures \cite{menushenkov}, where Lu-displacements by $ 0.2 - 0.3  $\AA{} are established.

\textit{Conclusion.} -- Lutetium dodecaboride has been studied here as the model rattling compound to clarify the nature of rattling modes and their influence on the crystal structure and properties. Precision measurements of the crystal structure and charge transport at low temperatures allow detecting the symmetry lowering, which may be attributed to coherent quasi-local vibrations (rattling modes) of Lu ions along singular direction in the dodecaboride lattice. It has also been shown that there is an extra source of lattice instability in LuB$_{12}$ related to the Jahn-Teller effect of B$_{12}$ clusters, which can manifest in concert with displacements of Lu atoms in oversized B$_{24}$ cages resulting in cooperative dynamic Jahn-Teller lattice distortions and conduction band changes.

\begin{acknowledgments}
This work was partially supported by grants from RFBR No. 15-02-02553, No. 16-02-00171 and by the Ministry of Education and Science of the Russian Federation within the framework of Increase Competitiveness Program of NUST «MISiS» (project No. К2-2017-023), implemented by governmental decree, N 211. The structure measurements were performed using the equipment of the Shared Research Center FSRC Crystallography and Photonics RAS. One of us (N.S.) acknowledges to V. Moshchalkov and F. Antson for useful discussions.
\end{acknowledgments}

\bibliography{paper_LuB12}

\providecommand{\noopsort}[1]{}\providecommand{\singleletter}[1]{#1}%
\begin{thebibliography}{25}%
\makeatletter
\providecommand \@ifxundefined [1]{%
 \@ifx{#1\undefined}
}%
\providecommand \@ifnum [1]{%
 \ifnum #1\expandafter \@firstoftwo
 \else \expandafter \@secondoftwo
 \fi
}%
\providecommand \@ifx [1]{%
 \ifx #1\expandafter \@firstoftwo
 \else \expandafter \@secondoftwo
 \fi
}%
\providecommand \natexlab [1]{#1}%
\providecommand \enquote  [1]{``#1''}%
\providecommand \bibnamefont  [1]{#1}%
\providecommand \bibfnamefont [1]{#1}%
\providecommand \citenamefont [1]{#1}%
\providecommand \href@noop [0]{\@secondoftwo}%
\providecommand \href [0]{\begingroup \@sanitize@url \@href}%
\providecommand \@href[1]{\@@startlink{#1}\@@href}%
\providecommand \@@href[1]{\endgroup#1\@@endlink}%
\providecommand \@sanitize@url [0]{\catcode `\\12\catcode `\$12\catcode
  `\&12\catcode `\#12\catcode `\^12\catcode `\_12\catcode `\%12\relax}%
\providecommand \@@startlink[1]{}%
\providecommand \@@endlink[0]{}%
\providecommand \url  [0]{\begingroup\@sanitize@url \@url }%
\providecommand \@url [1]{\endgroup\@href {#1}{\urlprefix }}%
\providecommand \urlprefix  [0]{URL }%
\providecommand \Eprint [0]{\href }%
\providecommand \doibase [0]{http://dx.doi.org/}%
\providecommand \selectlanguage [0]{\@gobble}%
\providecommand \bibinfo  [0]{\@secondoftwo}%
\providecommand \bibfield  [0]{\@secondoftwo}%
\providecommand \translation [1]{[#1]}%
\providecommand \BibitemOpen [0]{}%
\providecommand \bibitemStop [0]{}%
\providecommand \bibitemNoStop [0]{.\EOS\space}%
\providecommand \EOS [0]{\spacefactor3000\relax}%
\providecommand \BibitemShut  [1]{\csname bibitem#1\endcsname}%
\let\auto@bib@innerbib\@empty
\bibitem [{\citenamefont {Keppens}\ \emph {et~al.}(1998)\citenamefont
  {Keppens}, \citenamefont {Mandrus}, \citenamefont {Sales}, \citenamefont
  {Chakoumakos}, \citenamefont {Dai}, \citenamefont {Coldea}, \citenamefont
  {Maple}, \citenamefont {Gajewski}, \citenamefont {Freeman},\ and\
  \citenamefont {Bennington}}]{keppens98}%
  \BibitemOpen
  \bibfield  {author} {\bibinfo {author} {\bibfnamefont {V.}~\bibnamefont
  {Keppens}}, \bibinfo {author} {\bibfnamefont {D.}~\bibnamefont {Mandrus}},
  \bibinfo {author} {\bibfnamefont {B.}~\bibnamefont {Sales}}, \bibinfo
  {author} {\bibfnamefont {B.}~\bibnamefont {Chakoumakos}}, \bibinfo {author}
  {\bibfnamefont {P.}~\bibnamefont {Dai}}, \bibinfo {author} {\bibfnamefont
  {R.}~\bibnamefont {Coldea}}, \bibinfo {author} {\bibfnamefont
  {M.}~\bibnamefont {Maple}}, \bibinfo {author} {\bibfnamefont
  {D.}~\bibnamefont {Gajewski}}, \bibinfo {author} {\bibfnamefont
  {E.}~\bibnamefont {Freeman}}, \ and\ \bibinfo {author} {\bibfnamefont
  {S.}~\bibnamefont {Bennington}},\ }\href {\doibase 10.1038/27625} {\bibfield
  {journal} {\bibinfo  {journal} {Nature}\ }\textbf {\bibinfo {volume} {395}},\
  \bibinfo {pages} {876} (\bibinfo {year} {1998})}\BibitemShut {NoStop}%
\bibitem [{\citenamefont {Nakai}\ \emph {et~al.}(2008)\citenamefont {Nakai},
  \citenamefont {Ishida}, \citenamefont {Sugawara}, \citenamefont {Kikuchi},\
  and\ \citenamefont {Sato}}]{nakai2008}%
  \BibitemOpen
  \bibfield  {author} {\bibinfo {author} {\bibfnamefont {Y.}~\bibnamefont
  {Nakai}}, \bibinfo {author} {\bibfnamefont {K.}~\bibnamefont {Ishida}},
  \bibinfo {author} {\bibfnamefont {H.}~\bibnamefont {Sugawara}}, \bibinfo
  {author} {\bibfnamefont {D.}~\bibnamefont {Kikuchi}}, \ and\ \bibinfo
  {author} {\bibfnamefont {H.}~\bibnamefont {Sato}},\ }\href {\doibase
  10.1103/PhysRevB.77.041101} {\bibfield  {journal} {\bibinfo  {journal}
  {Phys.\ Rev. B}\ }\textbf {\bibinfo {volume} {77}},\ \bibinfo {pages}
  {041101} (\bibinfo {year} {2008})}\BibitemShut {NoStop}%
\bibitem [{\citenamefont {Yoshida}\ \emph {et~al.}(2007)\citenamefont
  {Yoshida}, \citenamefont {Arai}, \citenamefont {Kaido}, \citenamefont
  {Takigawa}, \citenamefont {Yonezawa}, \citenamefont {Muraoka},\ and\
  \citenamefont {Hiroi}}]{yoshida2007}%
  \BibitemOpen
  \bibfield  {author} {\bibinfo {author} {\bibfnamefont {M.}~\bibnamefont
  {Yoshida}}, \bibinfo {author} {\bibfnamefont {K.}~\bibnamefont {Arai}},
  \bibinfo {author} {\bibfnamefont {R.}~\bibnamefont {Kaido}}, \bibinfo
  {author} {\bibfnamefont {M.}~\bibnamefont {Takigawa}}, \bibinfo {author}
  {\bibfnamefont {S.}~\bibnamefont {Yonezawa}}, \bibinfo {author}
  {\bibfnamefont {Y.}~\bibnamefont {Muraoka}}, \ and\ \bibinfo {author}
  {\bibfnamefont {Z.}~\bibnamefont {Hiroi}},\ }\href {\doibase
  10.1103/PhysRevLett.98.197002} {\bibfield  {journal} {\bibinfo  {journal}
  {Phys.\ Rev. Lett.}\ }\textbf {\bibinfo {volume} {98}},\ \bibinfo {pages}
  {197002} (\bibinfo {year} {2007})}\BibitemShut {NoStop}%
\bibitem [{\citenamefont {Nolas}\ \emph {et~al.}(1998)\citenamefont {Nolas},
  \citenamefont {Cohn}, \citenamefont {Slack},\ and\ \citenamefont
  {Schujman}}]{nolas98}%
  \BibitemOpen
  \bibfield  {author} {\bibinfo {author} {\bibfnamefont {G.~S.}\ \bibnamefont
  {Nolas}}, \bibinfo {author} {\bibfnamefont {J.~L.}\ \bibnamefont {Cohn}},
  \bibinfo {author} {\bibfnamefont {G.~A.}\ \bibnamefont {Slack}}, \ and\
  \bibinfo {author} {\bibfnamefont {S.~B.}\ \bibnamefont {Schujman}},\ }\href
  {\doibase 10.1063/1.121747} {\bibfield  {journal} {\bibinfo  {journal} {Appl.
  Phys. Lett.}\ }\textbf {\bibinfo {volume} {73}},\ \bibinfo {pages} {178}
  (\bibinfo {year} {1998})}\BibitemShut {NoStop}%
\bibitem [{\citenamefont {Paschen}\ \emph {et~al.}(2014)\citenamefont
  {Paschen}, \citenamefont {Ikeda}, \citenamefont {Stefanoski},\ and\
  \citenamefont {Nolas}}]{paschen14}%
  \BibitemOpen
  \bibfield  {author} {\bibinfo {author} {\bibfnamefont {S.}~\bibnamefont
  {Paschen}}, \bibinfo {author} {\bibfnamefont {M.}~\bibnamefont {Ikeda}},
  \bibinfo {author} {\bibfnamefont {S.}~\bibnamefont {Stefanoski}}, \ and\
  \bibinfo {author} {\bibfnamefont {G.}~\bibnamefont {Nolas}},\ }in\ \href
  {\doibase 10.1007/978-94-017-9127-4_9} {\emph {\bibinfo {booktitle} {The
  Physics and Chemistry of Inorganic Clathrates}}},\ \bibinfo {series}
  {Springer Series in Materials Science}, Vol.\ \bibinfo {volume} {199},\
  \bibinfo {editor} {edited by\ \bibinfo {editor} {\bibfnamefont
  {G.}~\bibnamefont {Nolas}}}\ (\bibinfo  {publisher} {Springer Netherlands},\
  \bibinfo {year} {2014})\ Chap.~\bibinfo {chapter} {9}, pp.\ \bibinfo {pages}
  {249--276}\BibitemShut {NoStop}%
\bibitem [{\citenamefont {Asaki}\ \emph {et~al.}(2012)\citenamefont {Asaki},
  \citenamefont {Kotegawa}, \citenamefont {Tou}, \citenamefont {Onimaru},
  \citenamefont {Matsumoto}, \citenamefont {Inoue},\ and\ \citenamefont
  {Takabatake}}]{asaki12}%
  \BibitemOpen
  \bibfield  {author} {\bibinfo {author} {\bibfnamefont {K.}~\bibnamefont
  {Asaki}}, \bibinfo {author} {\bibfnamefont {H.}~\bibnamefont {Kotegawa}},
  \bibinfo {author} {\bibfnamefont {H.}~\bibnamefont {Tou}}, \bibinfo {author}
  {\bibfnamefont {T.}~\bibnamefont {Onimaru}}, \bibinfo {author} {\bibfnamefont
  {K.}~\bibnamefont {Matsumoto}}, \bibinfo {author} {\bibfnamefont
  {Y.}~\bibnamefont {Inoue}}, \ and\ \bibinfo {author} {\bibfnamefont
  {T.}~\bibnamefont {Takabatake}},\ }\href {\doibase 10.1143/JPSJ.81.023711}
  {\bibfield  {journal} {\bibinfo  {journal} {J. Phys. Soc. Jpn.}\ }\textbf
  {\bibinfo {volume} {81}},\ \bibinfo {pages} {023711} (\bibinfo {year}
  {2012})}\BibitemShut {NoStop}%
\bibitem [{\citenamefont {Akizuki}\ \emph {et~al.}(2015)\citenamefont
  {Akizuki}, \citenamefont {Yamada}, \citenamefont {Fujita}, \citenamefont
  {Taga}, \citenamefont {Kawakami}, \citenamefont {Mizumaki},\ and\
  \citenamefont {Tanaka}}]{akizuki15}%
  \BibitemOpen
  \bibfield  {author} {\bibinfo {author} {\bibfnamefont {Y.}~\bibnamefont
  {Akizuki}}, \bibinfo {author} {\bibfnamefont {I.}~\bibnamefont {Yamada}},
  \bibinfo {author} {\bibfnamefont {K.}~\bibnamefont {Fujita}}, \bibinfo
  {author} {\bibfnamefont {K.}~\bibnamefont {Taga}}, \bibinfo {author}
  {\bibfnamefont {T.}~\bibnamefont {Kawakami}}, \bibinfo {author}
  {\bibfnamefont {M.}~\bibnamefont {Mizumaki}}, \ and\ \bibinfo {author}
  {\bibfnamefont {K.}~\bibnamefont {Tanaka}},\ }\href {\doibase
  10.1002/anie.201504784} {\bibfield  {journal} {\bibinfo  {journal} {Angew.
  Chem. Int. Ed.}\ }\textbf {\bibinfo {volume} {54}},\ \bibinfo {pages} {10870}
  (\bibinfo {year} {2015})}\BibitemShut {NoStop}%
\bibitem [{\citenamefont {Sluchanko}\ \emph {et~al.}(2009)\citenamefont
  {Sluchanko}, \citenamefont {Bogach}, \citenamefont {Glushkov}, \citenamefont
  {Demishev}, \citenamefont {Lyubshov}, \citenamefont {Sluchanko},
  \citenamefont {Levchenko}, \citenamefont {Dukhnenko}, \citenamefont
  {Filipov}, \citenamefont {Gabani},\ and\ \citenamefont
  {Flachbart}}]{sluchanko09}%
  \BibitemOpen
  \bibfield  {author} {\bibinfo {author} {\bibfnamefont {N.~E.}\ \bibnamefont
  {Sluchanko}}, \bibinfo {author} {\bibfnamefont {A.~V.}\ \bibnamefont
  {Bogach}}, \bibinfo {author} {\bibfnamefont {V.~V.}\ \bibnamefont
  {Glushkov}}, \bibinfo {author} {\bibfnamefont {S.~V.}\ \bibnamefont
  {Demishev}}, \bibinfo {author} {\bibfnamefont {K.~S.}\ \bibnamefont
  {Lyubshov}}, \bibinfo {author} {\bibfnamefont {D.~N.}\ \bibnamefont
  {Sluchanko}}, \bibinfo {author} {\bibfnamefont {A.~V.}\ \bibnamefont
  {Levchenko}}, \bibinfo {author} {\bibfnamefont {A.~B.}\ \bibnamefont
  {Dukhnenko}}, \bibinfo {author} {\bibfnamefont {V.~B.}\ \bibnamefont
  {Filipov}}, \bibinfo {author} {\bibfnamefont {S.}~\bibnamefont {Gabani}}, \
  and\ \bibinfo {author} {\bibfnamefont {K.}~\bibnamefont {Flachbart}},\ }\href
  {\doibase 10.1134/S0021364009050099} {\bibfield  {journal} {\bibinfo
  {journal} {JETP Letters}\ }\textbf {\bibinfo {volume} {89}},\ \bibinfo
  {pages} {256} (\bibinfo {year} {2009})}\BibitemShut {NoStop}%
\bibitem [{\citenamefont {Alekseev}\ \emph {et~al.}(2014)\citenamefont
  {Alekseev}, \citenamefont {Nemkovski}, \citenamefont {Mignot}, \citenamefont
  {Clementyev}, \citenamefont {Ivanov}, \citenamefont {Rols}, \citenamefont
  {Bewley}, \citenamefont {Filipov},\ and\ \citenamefont
  {Shitsevalova}}]{alekseev14}%
  \BibitemOpen
  \bibfield  {author} {\bibinfo {author} {\bibfnamefont {P.~A.}\ \bibnamefont
  {Alekseev}}, \bibinfo {author} {\bibfnamefont {K.~S.}\ \bibnamefont
  {Nemkovski}}, \bibinfo {author} {\bibfnamefont {J.-M.}\ \bibnamefont
  {Mignot}}, \bibinfo {author} {\bibfnamefont {E.~S.}\ \bibnamefont
  {Clementyev}}, \bibinfo {author} {\bibfnamefont {A.~S.}\ \bibnamefont
  {Ivanov}}, \bibinfo {author} {\bibfnamefont {S.}~\bibnamefont {Rols}},
  \bibinfo {author} {\bibfnamefont {R.~I.}\ \bibnamefont {Bewley}}, \bibinfo
  {author} {\bibfnamefont {V.~B.}\ \bibnamefont {Filipov}}, \ and\ \bibinfo
  {author} {\bibfnamefont {N.~Y.}\ \bibnamefont {Shitsevalova}},\ }\href
  {\doibase 10.1103/PhysRevB.89.115121} {\bibfield  {journal} {\bibinfo
  {journal} {Phys.\ Rev. B}\ }\textbf {\bibinfo {volume} {89}},\ \bibinfo
  {pages} {115121} (\bibinfo {year} {2014})}\BibitemShut {NoStop}%
\bibitem [{\citenamefont {Alekseev}\ \emph {et~al.}(2008)\citenamefont
  {Alekseev}, \citenamefont {Grechnev}, \citenamefont {Shitsevalova},
  \citenamefont {Siemensmeyer}, \citenamefont {Sluchanko}, \citenamefont
  {Zogal},\ and\ \citenamefont {Flachbart}}]{alekseev08}%
  \BibitemOpen
  \bibfield  {author} {\bibinfo {author} {\bibfnamefont {P.}~\bibnamefont
  {Alekseev}}, \bibinfo {author} {\bibfnamefont {G.}~\bibnamefont {Grechnev}},
  \bibinfo {author} {\bibfnamefont {N.}~\bibnamefont {Shitsevalova}}, \bibinfo
  {author} {\bibfnamefont {K.}~\bibnamefont {Siemensmeyer}}, \bibinfo {author}
  {\bibfnamefont {N.}~\bibnamefont {Sluchanko}}, \bibinfo {author}
  {\bibfnamefont {O.}~\bibnamefont {Zogal}}, \ and\ \bibinfo {author}
  {\bibfnamefont {K.}~\bibnamefont {Flachbart}},\ }in\ \href@noop {} {\emph
  {\bibinfo {booktitle} {Rare Earths: Research and Applications}}},\ \bibinfo
  {editor} {edited by\ \bibinfo {editor} {\bibfnamefont {K.}~\bibnamefont
  {Delfrey}}}\ (\bibinfo  {publisher} {Nova},\ \bibinfo {address} {Commack, New
  York},\ \bibinfo {year} {2008})\ Chap.~\bibinfo {chapter} {2}, p.~\bibinfo
  {pages} {79}\BibitemShut {NoStop}%
\bibitem [{\citenamefont {Rybina}\ \emph {et~al.}(2010)\citenamefont {Rybina},
  \citenamefont {Nemkovski}, \citenamefont {Alekseev}, \citenamefont {Mignot},
  \citenamefont {Clementyev}, \citenamefont {Johnson}, \citenamefont {Capogna},
  \citenamefont {Dukhnenko}, \citenamefont {Lyashenko},\ and\ \citenamefont
  {Filippov}}]{rybina}%
  \BibitemOpen
  \bibfield  {author} {\bibinfo {author} {\bibfnamefont {A.~V.}\ \bibnamefont
  {Rybina}}, \bibinfo {author} {\bibfnamefont {K.~S.}\ \bibnamefont
  {Nemkovski}}, \bibinfo {author} {\bibfnamefont {P.~A.}\ \bibnamefont
  {Alekseev}}, \bibinfo {author} {\bibfnamefont {J.-M.}\ \bibnamefont
  {Mignot}}, \bibinfo {author} {\bibfnamefont {E.~S.}\ \bibnamefont
  {Clementyev}}, \bibinfo {author} {\bibfnamefont {M.}~\bibnamefont {Johnson}},
  \bibinfo {author} {\bibfnamefont {L.}~\bibnamefont {Capogna}}, \bibinfo
  {author} {\bibfnamefont {A.~V.}\ \bibnamefont {Dukhnenko}}, \bibinfo {author}
  {\bibfnamefont {A.~B.}\ \bibnamefont {Lyashenko}}, \ and\ \bibinfo {author}
  {\bibfnamefont {V.~B.}\ \bibnamefont {Filippov}},\ }\href {\doibase
  10.1103/PhysRevB.82.024302} {\bibfield  {journal} {\bibinfo  {journal}
  {Phys.\ Rev. B}\ }\textbf {\bibinfo {volume} {82}},\ \bibinfo {pages}
  {024302} (\bibinfo {year} {2010})}\BibitemShut {NoStop}%
\bibitem [{\citenamefont {Heinecke}\ \emph {et~al.}(1995)\citenamefont
  {Heinecke}, \citenamefont {Winzer}, \citenamefont {Noffke}, \citenamefont
  {Kranefeld}, \citenamefont {Grieb}, \citenamefont {Flachbart},\ and\
  \citenamefont {Paderno}}]{heinicke}%
  \BibitemOpen
  \bibfield  {author} {\bibinfo {author} {\bibfnamefont {M.}~\bibnamefont
  {Heinecke}}, \bibinfo {author} {\bibfnamefont {K.}~\bibnamefont {Winzer}},
  \bibinfo {author} {\bibfnamefont {J.}~\bibnamefont {Noffke}}, \bibinfo
  {author} {\bibfnamefont {H.}~\bibnamefont {Kranefeld}}, \bibinfo {author}
  {\bibfnamefont {H.}~\bibnamefont {Grieb}}, \bibinfo {author} {\bibfnamefont
  {K.}~\bibnamefont {Flachbart}}, \ and\ \bibinfo {author} {\bibfnamefont
  {Y.~B.}\ \bibnamefont {Paderno}},\ }\href {\doibase 10.1007/BF01324529}
  {\bibfield  {journal} {\bibinfo  {journal} {Z. Phys. B}\ }\textbf {\bibinfo
  {volume} {98}},\ \bibinfo {pages} {231} (\bibinfo {year} {1995})}\BibitemShut
  {NoStop}%
\bibitem [{\citenamefont {Jager}\ \emph {et~al.}(2006)\citenamefont {Jager},
  \citenamefont {Paluch}, \citenamefont {Zogał}, \citenamefont {Wolf},
  \citenamefont {Herzig}, \citenamefont {Filippov}, \citenamefont
  {Shitsevalova},\ and\ \citenamefont {Paderno}}]{jager}%
  \BibitemOpen
  \bibfield  {author} {\bibinfo {author} {\bibfnamefont {B.}~\bibnamefont
  {Jager}}, \bibinfo {author} {\bibfnamefont {S.}~\bibnamefont {Paluch}},
  \bibinfo {author} {\bibfnamefont {O.~J.}\ \bibnamefont {Zogał}}, \bibinfo
  {author} {\bibfnamefont {W.}~\bibnamefont {Wolf}}, \bibinfo {author}
  {\bibfnamefont {P.}~\bibnamefont {Herzig}}, \bibinfo {author} {\bibfnamefont
  {V.~B.}\ \bibnamefont {Filippov}}, \bibinfo {author} {\bibfnamefont {N.~Y.}\
  \bibnamefont {Shitsevalova}}, \ and\ \bibinfo {author} {\bibfnamefont
  {Y.~B.}\ \bibnamefont {Paderno}},\ }\href {\doibase
  10.1088/0953-8984/18/8/015} {\bibfield  {journal} {\bibinfo  {journal} {J.
  Phys. Condens. Matter}\ }\textbf {\bibinfo {volume} {18}},\ \bibinfo {pages}
  {2525} (\bibinfo {year} {2006})}\BibitemShut {NoStop}%
\bibitem [{\citenamefont {Mignot}\ \emph {et~al.}(2005)\citenamefont {Mignot},
  \citenamefont {Alekseev}, \citenamefont {Nemkovski}, \citenamefont
  {Regnault}, \citenamefont {Iga},\ and\ \citenamefont {Takabatake}}]{mignot}%
  \BibitemOpen
  \bibfield  {author} {\bibinfo {author} {\bibfnamefont {J.-M.}\ \bibnamefont
  {Mignot}}, \bibinfo {author} {\bibfnamefont {P.~A.}\ \bibnamefont
  {Alekseev}}, \bibinfo {author} {\bibfnamefont {K.~S.}\ \bibnamefont
  {Nemkovski}}, \bibinfo {author} {\bibfnamefont {L.-P.}\ \bibnamefont
  {Regnault}}, \bibinfo {author} {\bibfnamefont {F.}~\bibnamefont {Iga}}, \
  and\ \bibinfo {author} {\bibfnamefont {T.}~\bibnamefont {Takabatake}},\
  }\href {\doibase 10.1103/PhysRevLett.94.247204} {\bibfield  {journal}
  {\bibinfo  {journal} {Phys.\ Rev. Lett.}\ }\textbf {\bibinfo {volume} {94}},\
  \bibinfo {pages} {247204} (\bibinfo {year} {2005})}\BibitemShut {NoStop}%
\bibitem [{\citenamefont {Weng}\ \emph {et~al.}(2014)\citenamefont {Weng},
  \citenamefont {Zhao}, \citenamefont {Wang}, \citenamefont {Fang},\ and\
  \citenamefont {Dai}}]{weng}%
  \BibitemOpen
  \bibfield  {author} {\bibinfo {author} {\bibfnamefont {H.}~\bibnamefont
  {Weng}}, \bibinfo {author} {\bibfnamefont {J.}~\bibnamefont {Zhao}}, \bibinfo
  {author} {\bibfnamefont {Z.}~\bibnamefont {Wang}}, \bibinfo {author}
  {\bibfnamefont {Z.}~\bibnamefont {Fang}}, \ and\ \bibinfo {author}
  {\bibfnamefont {X.}~\bibnamefont {Dai}},\ }\href {\doibase
  10.1103/PhysRevLett.112.016403} {\bibfield  {journal} {\bibinfo  {journal}
  {Phys.\ Rev. Lett.}\ }\textbf {\bibinfo {volume} {112}},\ \bibinfo {pages}
  {016403} (\bibinfo {year} {2014})}\BibitemShut {NoStop}%
\bibitem [{\citenamefont {Sluchanko}\ \emph {et~al.}(2011)\citenamefont
  {Sluchanko}, \citenamefont {Azarevich}, \citenamefont {Bogach}, \citenamefont
  {Vlasov}, \citenamefont {Glushkov}, \citenamefont {Demishev}, \citenamefont
  {Maksimov}, \citenamefont {Tartakovskii}, \citenamefont {Filatov},
  \citenamefont {Flachbart}, \citenamefont {Gabani}, \citenamefont {Filippov},
  \citenamefont {Shitsevalova},\ and\ \citenamefont
  {Moshchalkov}}]{sluchanko11}%
  \BibitemOpen
  \bibfield  {author} {\bibinfo {author} {\bibfnamefont {N.}~\bibnamefont
  {Sluchanko}}, \bibinfo {author} {\bibfnamefont {A.}~\bibnamefont
  {Azarevich}}, \bibinfo {author} {\bibfnamefont {A.}~\bibnamefont {Bogach}},
  \bibinfo {author} {\bibfnamefont {I.}~\bibnamefont {Vlasov}}, \bibinfo
  {author} {\bibfnamefont {V.}~\bibnamefont {Glushkov}}, \bibinfo {author}
  {\bibfnamefont {S.}~\bibnamefont {Demishev}}, \bibinfo {author}
  {\bibfnamefont {A.}~\bibnamefont {Maksimov}}, \bibinfo {author}
  {\bibfnamefont {I.}~\bibnamefont {Tartakovskii}}, \bibinfo {author}
  {\bibfnamefont {E.}~\bibnamefont {Filatov}}, \bibinfo {author} {\bibfnamefont
  {K.}~\bibnamefont {Flachbart}}, \bibinfo {author} {\bibfnamefont
  {S.}~\bibnamefont {Gabani}}, \bibinfo {author} {\bibfnamefont
  {V.}~\bibnamefont {Filippov}}, \bibinfo {author} {\bibfnamefont
  {N.}~\bibnamefont {Shitsevalova}}, \ and\ \bibinfo {author} {\bibfnamefont
  {V.}~\bibnamefont {Moshchalkov}},\ }\href {\doibase
  10.1134/S1063776111080103} {\bibfield  {journal} {\bibinfo  {journal} {J.
  Exp. Theor. Phys.}\ }\textbf {\bibinfo {volume} {113}},\ \bibinfo {pages}
  {468} (\bibinfo {year} {2011})},\ \translation{JETP \textbf{140}, 536
  (2011)}\BibitemShut {NoStop}%
\bibitem [{\citenamefont {Bolotina}\ \emph {et~al.}(2016)\citenamefont
  {Bolotina}, \citenamefont {Verin}, \citenamefont {Shitsevalova},
  \citenamefont {Filippov},\ and\ \citenamefont {Sluchanko}}]{bolotina}%
  \BibitemOpen
  \bibfield  {author} {\bibinfo {author} {\bibfnamefont {N.~B.}\ \bibnamefont
  {Bolotina}}, \bibinfo {author} {\bibfnamefont {I.~A.}\ \bibnamefont {Verin}},
  \bibinfo {author} {\bibfnamefont {N.~Y.}\ \bibnamefont {Shitsevalova}},
  \bibinfo {author} {\bibfnamefont {V.~B.}\ \bibnamefont {Filippov}}, \ and\
  \bibinfo {author} {\bibfnamefont {N.~E.}\ \bibnamefont {Sluchanko}},\ }\href
  {\doibase 10.1134/S106377451602005X} {\bibfield  {journal} {\bibinfo
  {journal} {Crystallogr. Rep.}\ }\textbf {\bibinfo {volume} {61}},\ \bibinfo
  {pages} {181} (\bibinfo {year} {2016})}\BibitemShut {NoStop}%
\bibitem [{sup()}]{suppl}%
  \BibitemOpen
  \href@noop {} {\enquote {\bibinfo {title} {Supplementary materials},}\
  }\BibitemShut {NoStop}%
\bibitem [{\citenamefont {Sluchanko}\ \emph {et~al.}(2010)\citenamefont
  {Sluchanko}, \citenamefont {Azarevich}, \citenamefont {Bogach}, \citenamefont
  {Glushkov}, \citenamefont {Demishev}, \citenamefont {Kuznetsov},
  \citenamefont {Lyubshov}, \citenamefont {Filippov},\ and\ \citenamefont
  {Shitsevalova}}]{sluchanko10}%
  \BibitemOpen
  \bibfield  {author} {\bibinfo {author} {\bibfnamefont {N.}~\bibnamefont
  {Sluchanko}}, \bibinfo {author} {\bibfnamefont {A.}~\bibnamefont
  {Azarevich}}, \bibinfo {author} {\bibfnamefont {A.}~\bibnamefont {Bogach}},
  \bibinfo {author} {\bibfnamefont {V.}~\bibnamefont {Glushkov}}, \bibinfo
  {author} {\bibfnamefont {S.}~\bibnamefont {Demishev}}, \bibinfo {author}
  {\bibfnamefont {A.~V.}\ \bibnamefont {Kuznetsov}}, \bibinfo {author}
  {\bibfnamefont {K.~S.}\ \bibnamefont {Lyubshov}}, \bibinfo {author}
  {\bibfnamefont {V.}~\bibnamefont {Filippov}}, \ and\ \bibinfo {author}
  {\bibfnamefont {N.}~\bibnamefont {Shitsevalova}},\ }\href {\doibase
  10.1134/S1063776110080212} {\bibfield  {journal} {\bibinfo  {journal} {J.
  Exp. Theor. Phys.}\ }\textbf {\bibinfo {volume} {111}},\ \bibinfo {pages}
  {279} (\bibinfo {year} {2010})},\ \translation{JETP \textbf{138}, 315
  (2010)}\BibitemShut {NoStop}%
\bibitem [{\citenamefont {Sluchanko}\ \emph {et~al.}(2016)\citenamefont
  {Sluchanko}, \citenamefont {Azarevich}, \citenamefont {Anisimov},
  \citenamefont {Bogach}, \citenamefont {Gavrilkin}, \citenamefont {Gilmanov},
  \citenamefont {Glushkov}, \citenamefont {Demishev}, \citenamefont
  {Khoroshilov}, \citenamefont {Dukhnenko}, \citenamefont {Mitsen},
  \citenamefont {Shitsevalova}, \citenamefont {Filipov}, \citenamefont
  {Voronov},\ and\ \citenamefont {Flachbart}}]{sluchanko16}%
  \BibitemOpen
  \bibfield  {author} {\bibinfo {author} {\bibfnamefont {N.}~\bibnamefont
  {Sluchanko}}, \bibinfo {author} {\bibfnamefont {A.}~\bibnamefont
  {Azarevich}}, \bibinfo {author} {\bibfnamefont {M.}~\bibnamefont {Anisimov}},
  \bibinfo {author} {\bibfnamefont {A.}~\bibnamefont {Bogach}}, \bibinfo
  {author} {\bibfnamefont {S.}~\bibnamefont {Gavrilkin}}, \bibinfo {author}
  {\bibfnamefont {M.}~\bibnamefont {Gilmanov}}, \bibinfo {author}
  {\bibfnamefont {V.}~\bibnamefont {Glushkov}}, \bibinfo {author}
  {\bibfnamefont {S.}~\bibnamefont {Demishev}}, \bibinfo {author}
  {\bibfnamefont {A.}~\bibnamefont {Khoroshilov}}, \bibinfo {author}
  {\bibfnamefont {A.}~\bibnamefont {Dukhnenko}}, \bibinfo {author}
  {\bibfnamefont {K.}~\bibnamefont {Mitsen}}, \bibinfo {author} {\bibfnamefont
  {N.}~\bibnamefont {Shitsevalova}}, \bibinfo {author} {\bibfnamefont
  {V.}~\bibnamefont {Filipov}}, \bibinfo {author} {\bibfnamefont
  {V.}~\bibnamefont {Voronov}}, \ and\ \bibinfo {author} {\bibfnamefont
  {K.}~\bibnamefont {Flachbart}},\ }\href {\doibase 10.1103/PhysRevB.93.085130}
  {\bibfield  {journal} {\bibinfo  {journal} {Phys.\ Rev. B}\ }\textbf
  {\bibinfo {volume} {93}},\ \bibinfo {pages} {085130} (\bibinfo {year}
  {2016})}\BibitemShut {NoStop}%
\bibitem [{\citenamefont {Franz}\ and\ \citenamefont {Werheit}(1989)}]{franz}%
  \BibitemOpen
  \bibfield  {author} {\bibinfo {author} {\bibfnamefont {R.}~\bibnamefont
  {Franz}}\ and\ \bibinfo {author} {\bibfnamefont {H.}~\bibnamefont
  {Werheit}},\ }\href {http://stacks.iop.org/0295-5075/9/i=2/a=009} {\bibfield
  {journal} {\bibinfo  {journal} {EPL (Europhysics Letters)}\ }\textbf
  {\bibinfo {volume} {9}},\ \bibinfo {pages} {145} (\bibinfo {year}
  {1989})}\BibitemShut {NoStop}%
\bibitem [{\citenamefont {Bersuker}\ and\ \citenamefont
  {Polinger}(1989)}]{bersuker}%
  \BibitemOpen
  \bibfield  {author} {\bibinfo {author} {\bibfnamefont {I.~B.}\ \bibnamefont
  {Bersuker}}\ and\ \bibinfo {author} {\bibfnamefont {V.~Z.}\ \bibnamefont
  {Polinger}},\ }\href@noop {} {\emph {\bibinfo {title} {Vibronic Interactions
  in Molecules and Crystals}}}\ (\bibinfo  {publisher} {Springer},\ \bibinfo
  {address} {Berlin},\ \bibinfo {year} {1989})\BibitemShut {NoStop}%
\bibitem [{\citenamefont {Reznik}(2012)}]{REZNIK201275}%
  \BibitemOpen
  \bibfield  {author} {\bibinfo {author} {\bibfnamefont {D.}~\bibnamefont
  {Reznik}},\ }\href {\doibase http://dx.doi.org/10.1016/j.physc.2012.01.024}
  {\bibfield  {journal} {\bibinfo  {journal} {Physica C: Superconductivity}\
  }\textbf {\bibinfo {volume} {481}},\ \bibinfo {pages} {75 } (\bibinfo {year}
  {2012})}\BibitemShut {NoStop}%
\bibitem [{\citenamefont {Sluchanko}\ \emph {et~al.}(2014)\citenamefont
  {Sluchanko}, \citenamefont {Azarevich}, \citenamefont {Gavrilkin},
  \citenamefont {Glushkov}, \citenamefont {Demishev}, \citenamefont
  {Shitsevalova},\ and\ \citenamefont {Filipov}}]{sluchanko14}%
  \BibitemOpen
  \bibfield  {author} {\bibinfo {author} {\bibfnamefont {N.~E.}\ \bibnamefont
  {Sluchanko}}, \bibinfo {author} {\bibfnamefont {A.~N.}\ \bibnamefont
  {Azarevich}}, \bibinfo {author} {\bibfnamefont {S.}~\bibnamefont
  {Gavrilkin}}, \bibinfo {author} {\bibfnamefont {V.~V.}\ \bibnamefont
  {Glushkov}}, \bibinfo {author} {\bibfnamefont {S.~V.}\ \bibnamefont
  {Demishev}}, \bibinfo {author} {\bibfnamefont {N.~Y.}\ \bibnamefont
  {Shitsevalova}}, \ and\ \bibinfo {author} {\bibfnamefont {V.~B.}\
  \bibnamefont {Filipov}},\ }\href {\doibase 10.1134/S0021364013220141}
  {\bibfield  {journal} {\bibinfo  {journal} {JETP Letters}\ }\textbf {\bibinfo
  {volume} {98}},\ \bibinfo {pages} {578} (\bibinfo {year} {2014})}\BibitemShut
  {NoStop}%
\bibitem [{\citenamefont {Menushenkov}\ \emph {et~al.}(2013)\citenamefont
  {Menushenkov}, \citenamefont {Yaroslavtsev}, \citenamefont {Zaluzhnyy},
  \citenamefont {Kuznetsov}, \citenamefont {Chernikov}, \citenamefont
  {Shitsevalova},\ and\ \citenamefont {Filippov}}]{menushenkov}%
  \BibitemOpen
  \bibfield  {author} {\bibinfo {author} {\bibfnamefont {A.~P.}\ \bibnamefont
  {Menushenkov}}, \bibinfo {author} {\bibfnamefont {A.~A.}\ \bibnamefont
  {Yaroslavtsev}}, \bibinfo {author} {\bibfnamefont {I.~A.}\ \bibnamefont
  {Zaluzhnyy}}, \bibinfo {author} {\bibfnamefont {A.~V.}\ \bibnamefont
  {Kuznetsov}}, \bibinfo {author} {\bibfnamefont {R.~V.}\ \bibnamefont
  {Chernikov}}, \bibinfo {author} {\bibfnamefont {N.~Y.}\ \bibnamefont
  {Shitsevalova}}, \ and\ \bibinfo {author} {\bibfnamefont {V.~B.}\
  \bibnamefont {Filippov}},\ }\href {\doibase 10.1134/S002136401316011X}
  {\bibfield  {journal} {\bibinfo  {journal} {JETP Letters}\ }\textbf {\bibinfo
  {volume} {98}},\ \bibinfo {pages} {165} (\bibinfo {year} {2013})}\BibitemShut
  {NoStop}%
\end{thebibliography}%

\end{document}